# Three years later: gender differences in the advisor's impact on career choices in astronomy and astrophysics

Rachel Ivie, Susan White, and Raymond Y. Chu*
*American Institute of Physics, One Physics Ellipse, College Park, Maryland, 20740, USA

The Longitudinal Study of Astronomy Graduate Students (LSAGS) arose from the 2003 Women in Astronomy Conference, where it was noted that a majority of young members of the American Astronomical Society were women. The astronomy community wishes to make every effort to retain young women in astronomy, so they commissioned a longitudinal study to be conducted that would pinpoint the factors that contribute to retention in general, with a focus on differences between women and men. The LSAGS follows a cohort of people who were graduate students in astronomy or astrophysics during 2006-07. The first survey was conducted during 2007-08, the second during 2012-13, and the third during 2015. The analysis presented in this paper, which is an update to our previous paper on this topic, used a subset of the respondents, all of whom had PhDs in astronomy, astrophysics, or a related field at the time of the third survey. We tested the effects of four major concepts on attrition from physics and astronomy. These concepts included: the imposter syndrome, mentoring and advising during graduate school, the so-called "two-body problem" that occurs when a couple needs to find two jobs in the same geographic area, and gender of the respondent. Having a mentor in grad school did not contribute to working outside of physics or astronomy. Showing characteristics of the imposter syndrome and gender of the respondent had indirect effects on working outside the field. Encouragement of the graduate advisor, the two-body problem, and completing a postdoc, had significant direct effects on working in physics or astronomy. This research identifies specific areas of concern that can be addressed by the scientific community to increase the retention of all people, but especially women, in astronomy and astrophysics.

## I. INTRODUCTION

There is evidence that women, once they have completed doctoral degrees in physical sciences, advance up the academic ladder at about the same rates as men [1, 2, 3]. However, it is still true that many women are lost to physical sciences at some point along the way. In 2007, the National Academies reviewed the literature examining issues for women in science and engineering; the study concluded that women are lost to science and engineering careers at every educational transition for several reasons, including documented discrimination, implicit bias, evaluation criteria that disadvantage women, and the structure of academic organizations [4]. In physics, about 50% of high school students are female [5], but the proportion of bachelor's degrees earned by women is only 20% [6]. The attrition is similar in astronomy and astrophysics, although women earn just over one-third of the bachelor's degrees [7].

In 2003, the Committee for the Status of Women in Astronomy (CSWA) and the AAS Council concluded that a longitudinal study was needed to collect data about variables that affect career choices in astronomy and to determine whether any of these variables affect men and women differently. The resulting study, the Longitudinal Study of Astronomy Graduate Students (LSAGS), is a joint project of the American Astronomical Society (AAS) and the American Institute of Physics (AIP).

This paper provides updates to previously published work [8] using responses from three rounds of the LSAGS. We examine the issue of attrition from astronomy and astrophysics for men and women who have earned doctorates and were not in a postdoc at the time of the 2015 survey. Using a cohort of people who had been graduate students in astronomy or astrophysics during 2006-2007, we obtained data at three points in time: 1) during 2007-8, 2) during 2012-13, and 3) during 2015. Out of 2056 graduate students contacted in 2007-08, we had 1143 usable responses to the first survey.  837 responded to the second survey, and 814 responded to the third survey. 465 responded to all three rounds. We are grateful to the respondents for completing the survey.

We present an analysis that shows whether respondents were still working in the field at the time of the third survey and which variables are likely to influence that outcome. Our focus is on whether gender of the respondents contributes to attrition, controlling for other factors that may influence working in or out of field.  In our analysis of data from the first and second rounds, we found that gender did not directly affect attrition. Respondents' opinions of advisors, along with measures of work-family balance, were among the factors that explained attrition. Since the gender of the respondent predicts the factors that affect attrition, we found that gender has an indirect effect on attrition.

## II. BACKGROUND

### A. Attrition in Science Careers

Xie and Shauman looked at gender differences in the careers of scientists. They found that women with master's degrees in science and engineering were less likely than men to work in science and engineering jobs. Women's family status was a major determinant of their pursuit of science and engineering occupations after graduate training in these fields [9]. Because of the nature of their data, Xie and Shauman were not able to show results for specific fields such as astronomy and physics.

An analysis of data from the NSF's *Survey of Doctoral Recipients* (SDR) showed that men were more likely than women to leave academe for non-academic jobs, but did not provide data on the propensity of people to leave their specific fields altogether for another field [4]. In addition, results from the SDR are almost always reported by broad field, such as "physical science," which means that trends in smaller fields such as physics are astronomy are lost among trends for larger fields that have much different economic prospects.

The fact that national data relies on samples such as the SDR means that generally, not enough information is available to draw conclusions about astronomy, which has even fewer people than physics. However, working environments and job prospects differ greatly by specific field within physical sciences (consider, for example, the differences between chemistry and astronomy). Therefore, to understand factors causing attrition, it is essential to study specific fields so that we can determine factors contributing to individual career decisions.

One study that focused specifically on physicists' careers is Joseph Hermanowicz's examination of the careers of 55 physicists at various types of universities [10]. This study may be the only longitudinal study of physicists' careers. Hermanowicz interviewed 55 physics faculty members once, and then followed up 10 years later as several in his study were beginning to retire. Hermanowicz finds that the characteristics of the college or university in which the physicists worked strongly influence career outcomes.

### B. Attrition from astronomy

According to The National Research Council's 2010 Decadal Survey of Astronomy and Astrophysics, graduate training in astronomy often emphasizes academic careers for students. However, the survey committee recognized the broad applicability of astronomy training to other fields and recommended that professional training should correspond to the actual range of career paths taken by those who have received graduate training in astronomy and astrophysics. The committee estimated that at least 20% of PhD astronomers leave the field at some point after earning their doctorates [11]. In the Decadal Survey's report, the percentage of astronomers who leave the field before receiving a PhD was not addressed because that number is unknown.

In our previous work [8], we found that changing advisors, limiting career options for someone else, relocating for a spouse or partner, and not doing a postdoc directly contributed to working out of field in 2012. In addition, advisor rating and imposter score had indirect effects. Gender also had an indirect on working out of field because it affected the imposter score, the advisor rating, and whether one relocated for a spouse or partner. These effects are depicted in Figure 1. Specifically, the direct effects on working out of field (depicted by blue arrows in Figure 1) are

- Respondents who changed advisors in graduate school were more likely to be working out of astronomy in 2012,
- Respondents who limited their career options for someone else were more likely to be working out of astronomy in 2012,
- Respondents who had relocated for a spouse or partner were more likely to be working out of astronomy in 2012, and
- Respondents who had not taken a postdoc were more likely to be working out of astronomy in 2012.

The indirect effects on working out of field (depicted by red arrows in Figure 1) are

- Respondents with a higher imposter score were more likely to change advisors,
- Respondents with a higher advisor rating were less likely to change advisors and less likely to limit their career options for someone else,
- Respondents who relocated for a spouse or partner were more likely to say they had limited their career options for someone else, and
- Women were more likely to have a higher imposter score and a lower advisor rating and were more likely to have relocated for a spouse or partner.

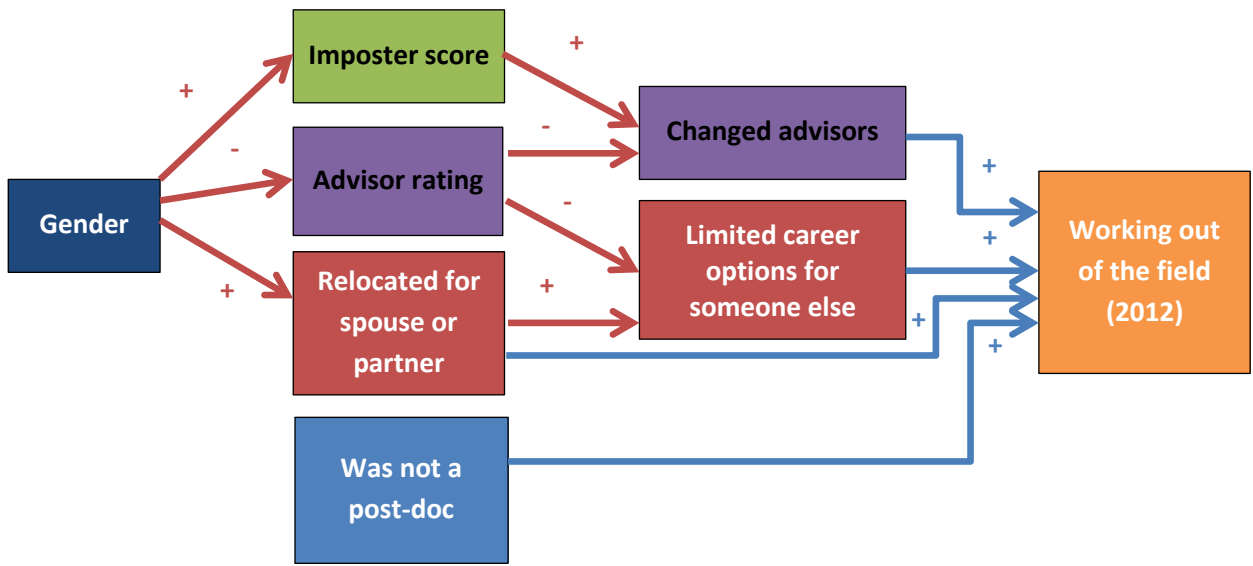

Figure 1: Direct and Indirect effects on working outside astronomy in 2012

## C. Factors Affecting Attrition

### 1. Imposter syndrome

The first questionnaire used on the LSAGS measured characteristics of the imposter syndrome because we hypothesized that students with the imposter syndrome would be more likely to leave the field. The imposter syndrome was first used by psychologists Pauline Clance and Suzanne Imes in 1978 [12] to describe highly successful women who nevertheless had difficulty internally recognizing their own achievements and continued to feel as though they were imposters in their careers. Since that time, additional research demonstrated that men can also exhibit characteristics of the imposter syndrome. In further describing the imposter syndrome, Langford and Clance [13] wrote that the syndrome is defined by "believing that one's accomplishments came about not through genuine ability, but as a result of having been lucky, having worked harder than others, and having manipulated other people's impressions." One key aspect of the imposter syndrome is the attribution of success to factors beyond individual control, such as luck, while attributing the success of others to skill or knowledge. But it is not just external factors to which those with the imposter syndrome attribute their successes. People with the imposter syndrome can also discount their successes by attributing them to hard work, while believing that others sail through based on natural talent. The imposter syndrome also can cause individuals to believe that people will soon realize that they are not really capable after all [13]. In the first survey in the LSAGS, women showed characteristics consistent with the imposter syndrome [14].

### 2. Advisor's Attitudes and Encouragement

In our second survey, we included questions about mentors because mentoring is often cited as a mechanism to improve the retention of students, particularly women [15]. Our 2012 data did show that women were much more likely than men to have had a mentor other than their advisor while in graduate school (62% to 48%). It could be that women are more likely to seek out a mentor because their relationships with their advisors were not as supportive as those the men had. Schlosser *et al*. [16] suggest that an advising relationship may be positive, neutral, or negative, while mentoring refers to "an inherently positive relationship." That could help explain why we found that having a mentor other than one's advisor while in graduate school did not affect whether one was working in astronomy or physics after earning a PhD and completing any postdocs. Women were more likely to have a mentor; while men, on average, had better relationships with their advisors. So, the impact of having a mentor was possibly offset by the relationship with one's advisor.

Research has examined the role of the advisor in attrition from graduate programs. (See, for example, Lovitts [17] and Golde [18].) Weidman and Stein [19] and Gardner [20] have considered the role of the academic advisor in socialization in the discipline. Other researchers have examined the effect of one's advisor on career commitment. Litzler, Lange, and Brainard [21] find that students are more likely to respond positively to the question "to what extent has your academic experience in your department reaffirmed your career choice?" if they feel they have a good relationship with their advisor. We found very similar results. Zhao, Golde and McCormick [22] examined factors affecting advisor choice and found three: advisor reputation, intellectual compatibility, and pragmatic benefit. They examined factors that might affect each of these. The factors included gender, marital status, race, whether the respondent had children, the parents' highest degree, the age of the respondent, and the academic discipline. The only instance in which gender was significant was pragmatic benefit, and the *p*-value was between 0.01 and 0.05. Pragmatic benefit includes the availability of funding from the advisor to support the student, how interesting the student finds the research, and how much the student perceives the advisor to foster an attractive working environment. Across multiple disciplines, women were slightly more likely to consider pragmatic benefit when choosing an advisor.

*3. Two-body problem*

Increasingly, universities are hiring dual-career couples [23]. In physics, these dual-career couples are often said to have the "two-body" problem. In 1998, physicists McNeil and Sher defined the "two-body" problem as "the difficulty of finding two professional jobs (possibly two physics jobs) in the same geographic location." Because women are more likely to be married to other academics than men are, women may be more likely to experience the two-body problem [24]. A more recent study shows that the number one reason women turned down offers of employment at academic institutions is because their partners did not find appropriate employment at the new location [23]. McNeil and Sher hypothesized that the frustration of finding jobs may ultimately lead one of the partners in a dual-career couple to leave the field altogether [24]. So far, studies of dual career couples have collected data from people still employed in academics, so the effects of the "two-body" problem on leaving science are unknown.

*4. Being female*

Generally, the studies of women's careers in science have not been able to include data on women in specific fields such as astronomy. Nevertheless, there have been studies of the representation of women in astronomy, several of which attempted to look for drop-out points for women. These studies have not necessarily presented a consistent picture. For example, Hoffman & Urry [25] concluded in their 2004 analysis of three Space Telescope Science Institute surveys in 1992, 1999, and 2003, that while women were progressing at about the same rate as men during the 1990s, differential attrition may have been occurring between 1996 and 2003. An AIP report published in 2005 by Ivie & Ray [3] concluded that there appeared to be no leak in the pipeline at the faculty level for either physics or astronomy, although there may have been a small leak in astronomy between bachelor's and PhDs. Yet this study revealed a dramatic leak from high school to college physics. In a re-analysis of AIP's data, Bagenal [26] concluded a significant differential leak remains for women in astronomy from undergraduate to graduate school, but that the percentage of women within the three main professorial ranks was approximately what was expected considering the number of PhDs awarded to women. Marvel [27] reported that snapshot surveys of AAS membership revealed dramatic changes in the demographics of the AAS, with women making up 60% of the youngest AAS members in 2004. The question raised by this is whether these younger women will stay in the field.

III. ABOUT THE DATA

The Longitudinal Study of Astronomy Graduate Students arose from the recommendations discussed at the *Women in Astronomy II* conference held in Pasadena, CA in 2003. One of the key recommendations was that the AAS commission a "longitudinal study of young women in astronomy," in order to "measure whether there is differential attrition of women from the pipeline and if so, to learn the reasons for it. . . ." [28] The Committee for the Status of Women in Astronomy and the AAS Office of Education organized the LSAGS in 2006 and established a project team to conduct the study. We have sent three questionnaires to potential respondents who were graduate students in astronomy or astrophysics in 2006-07.

A. First survey

In 2007, we identified 2056 possible astronomy and astrophysics graduate students from the AAS junior membership and American Institute of Physics survey data. The first LSAGS survey was carried out in 2007-08; 1143 individuals (447 women, 696 men) responded to the first survey. The questionnaire instrument was available both on the web and on paper; the first questionnaire can be accessed using http://aipsurveys.aip.org/cgi-bin/astrograd2006.pl. We contacted respondents both via e-mail and postal mail to increase the likelihood that everyone received a survey invitation. We used four e-mail contacts and three paper contacts. This study is a cohort study. We have collected data from the same individuals at more than one point in time, and every respondent to all three surveys was a graduate student in astronomy or astrophysics during the 2006-07 academic year.

B. Second survey

Data collection for the second round spanned 2012-13. Where possible, we updated the contact information for the cohort using 1) contact information provided by the respondents to the first survey, 2) AAS membership lists, and 3) a postal address updater service. At this time, 1555 individuals remained in the cohort (we discovered during the first survey that not all of the 2056 people in the cohort were astronomy graduate students during 2006-07, and we discovered we did not have current contact information for others). We sent three e-mail requests and a final request via by postal mail. This survey, which had multiple skip patterns so that questions were tailored to the specific situations of respondents, was available only on the web. The second questionnaire can be accessed using http://aipsurveys.aip.org/cgi-bin/astrograd2012.pl.  There were complicated skip patterns in this questionnaire; interested researchers should go through it several times using different responses to see all the skips. The key questions are the last question on each page. The postal mail request gave respondents a url directing them to the survey. We received 837 responses, and of these, 666 also responded to the first survey.  Unfortunately, time constraints meant that we could not collect as many respondents to the second survey as we did to the first.

C. Third Survey

Data collection for the third survey spanned 2015 and 2016. As we did for the second round, we made every possible effort to update the contact information. We sent four e-mail requests between November 16, 2015, and February 23, 2016. This survey also had multiple skip patterns so that questions were tailored to the specific situations of the respondents. The questionnaire can be accessed using https://aipsurveys.aip.org/cgi-bin/astrogradphd2015.pl. We received 814 responses to the third survey. 465 people responded to all three surveys.

IV. MEASUREMENT

A. Attrition

We used the respondents' fields of employment to determine whether or not they had left the fields of astronomy, astrophysics, and physics. We decided to include respondents who reported being employed in physics as employed in field. Our determination about what is in or out of field does not necessarily match the respondents' perceptions. For example, it was not uncommon for respondents employed in planetary science to report that they were working out of field, although we considered them in field. In the model of this dependent variable, our analysis was limited to those who were not currently postdocs and who had answered all the questions in the model. We had about 240 useable responses; about 30% were working outside the fields of astronomy or physics.

### B. Factors that may influence attrition

#### 1. Imposter Syndrome

Respondents to the first survey answered a set of 7 questions designed to measure characteristics of the imposter syndrome. Answers to these questions were combined into a score ranging from 7-35, with a higher score indicating respondents who feel more like imposters in their field.

#### 2. Advisor's Attitude and Encouragement

On the second survey, we asked respondents a yes/no question about whether they had a mentor other than their advisors during graduate school. We asked a yes/no question about whether they had changed advisors during graduate school, which could indicate a level of dissatisfaction at least with their first advisor.

We also asked several questions about respondents' relationships with their advisors. These four questions asked respondents to indicate on a four-point scale whether their advisors were helpful, encouraging, easy to discuss ideas with, and gave adequate input. The responses were summed into a score ranging from 4-16, with a higher score indicating a higher level of satisfaction with the respondents' advisors.

On the third survey, we added sixteen questions about advisors to allow us to explore the relationship between one's advisor and persistence in astronomy. The questions covered ways the advisor supported the student in research, in growing as a professional, and in more general ways. Many of the advisor questions we used came from the 2013 ACS Graduate Student Survey [29] conducted by the American Chemical Society. In their report, the authors of the ACS study note that "Men, more than women, reported that their advisors engaged in behaviors that help them advance professionally." (p. 5) Our findings are consistent with these.

#### 3. Two-body problem

We asked respondents three yes/no questions designed to measure whether they had experienced a situation related to the need to find two jobs in the same geographic area:

1) Have they relocated because of a spouse or partner?
2) Do they maintain a residence in a different location from their family in order to work or study?
3) Have they limited their career options because of someone else?

Each of these was included separately (rather than as a score) in the models tested.

#### 4. Other variables

Because they may influence attrition, we also included a binary measure of gender (male/female), postdoc status at the time of the second survey (currently a post doc, completed a postdoc, and a PhD who has never been a postdoc), and the number of years that have elapsed since the respondents earned their PhDs. We hypothesized that people who have had postdocs may be less likely to leave astronomy, and we wanted to control for differences in the amount of time that respondents had been out of graduate school.

Among the respondents to the third survey, 90% had completed PhDs in astronomy, astrophysics or a related field. Of these, about 28% were current postdocs, 55% had completed one or more postdocs, and 17% had never been postdocs. All the respondents included in this analysis had PhDs and were not currently a postdoc at the time of the third survey.

## V. ANALYSIS AND RESULTS

### A. Bivariate gender differences

The advisor questions asked both on the ACS questionnaire and our longitudinal study are shown in Table 1. Respondents were asked to indicate their level of agreement with each of the following statements. In all cases in which there was a statistically significant gender difference, it was men who were more likely than women to have a positive evaluation of their advisors. Then statements on which men were more likely to agree in the LSAGS study are shown in bold.

| My advisor … | Are men more likely to agree? | |
|---|---|---|
| | LSAGS Results | ACS Results* |
| **Encouraged me to attain my goals**** | Yes, $p < 0.01$ | Yes, $p < 0.01$ |
| **Advocated for me** | Yes, $p < 0.01$ | Yes, $0.05 > p > 0.01$ |
| **Supported my career path of choice** | Yes, $p < 0.01$ | Yes, $0.05 > p > 0.01$ |
| **Gave the appropriate level of credit to me for my research contributions** | Yes, $0.05 > p > 0.01$ | Yes, $p < 0.01$ |
| Encouraged me to present our research at scientific conferences | No statistically significant difference | Yes, $p < 0.01$ |
| Gave regular feedback on my research | No statistically significant difference | Yes, $p < 0.01$ |
| Engaged me in writing grant proposals | No statistically significant difference | Yes, $p < 0.01$ |
| Helped me to develop professional relationships | No statistically significant difference | Yes, $0.05 > p > 0.01$ |
| Provided information about academic career paths | No statistically significant difference | Yes, $0.10 > p > 0.05$ |
| Provided information about non-academic career paths | No statistically significant difference | No statistically significant difference |
| Modeled good professional relationships | No statistically significant difference | No statistically significant difference |
| Took time to learn about my background, interests, and/or personal relationships | No statistically significant difference | No statistically significant difference |

**Table 1: Respondents' Evaluations of Advisors**

\* There were more participants in the ACS study than in LSAGS. Thus, their tests had more statistical power. Therefore, it is not surprising that were a larger number of statistically significant differences between the responses of men and women in the ACS study.

\*\* Current astronomy students saw present-tense versions of these questions. We show past-tense versions here since that is what degree recipients saw and what is used in the model.

In Figure 2, we display the responses from LSAGS participants for the first statement in Table 1. Almost 95% of the men either agreed or strongly agreed with the statement; this was true for just over 80% of the women. Women were also three times more likely than men to disagree or disagree strongly (6% of men and 18% of women). For the other questions where the responses were statistically significantly different, they followed patterns like that in Figure 2.

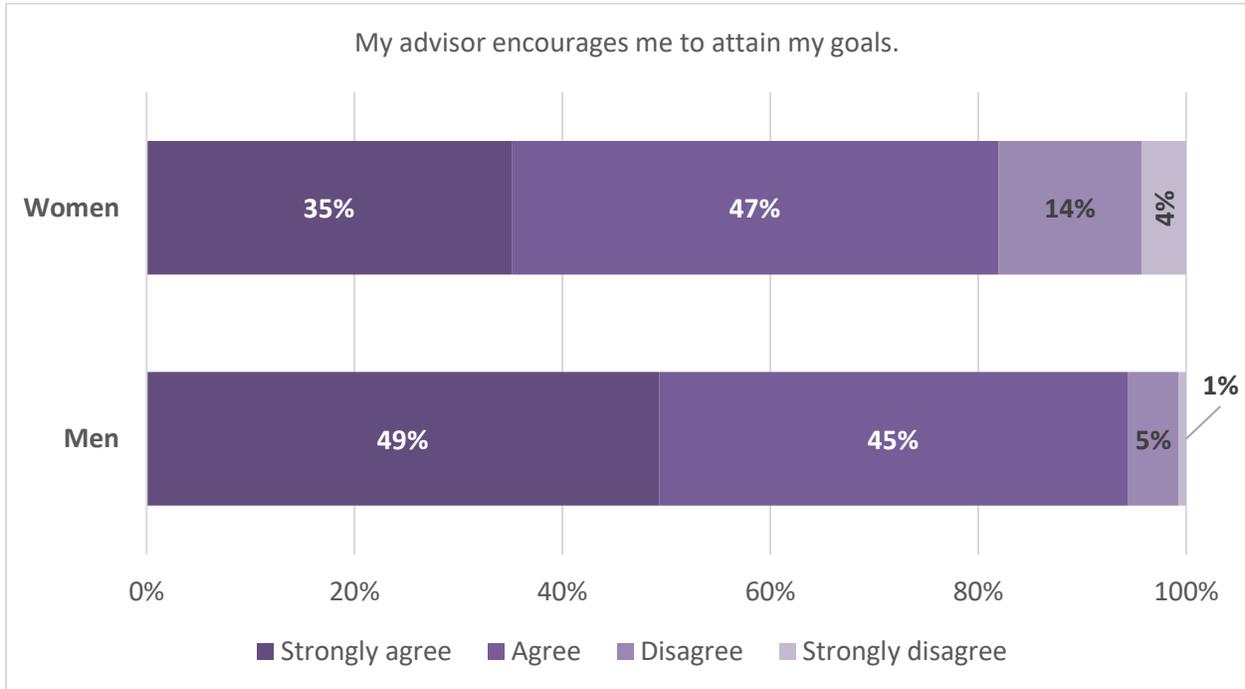

**Figure 2: Men's and women's responses to "My advisor encourages me to attain my goals."**

Since we wanted to examine differences in attrition by gender, we first looked for gender differences in each factor that could influence attrition. Overall, 39% of the respondents included in the analysis were women. For five measures, we found highly significant differences ($p$-value < 0.01) by gender.

- Women were much more likely to have relocated for a spouse or partner than men.
- Women were much more likely to maintain a separate residence for work or study than men.
- Women had a lower opinion of their advisors than men.
- Women were much more likely than men to have had a mentor other than their advisor during graduate school.
- Women were more likely to feel like imposters than men.

Table 2 provides a summary of the variables included in the analysis with more details on differences between men and women on these variables.

| Variable | Overall | Differences by gender? | Level of Significance* |
|---|---|---|---|
| Relocated for spouse or partner | 24% | Women (33%) were much more likely to have relocated for a spouse or partner than men (18%). | Highly significant |
| Maintained residence in different location from family in order to work or study | 13% | Women (21%) were much more likely to maintain a separate residence than men (9%). | Highly significant |
| Advisor rating (scale is 4 to 16 with a lower score implying a worse opinion of the advisor) | 13.6 | Women's scores (13.2) were lower than men's (13.9) meaning women had a lower opinion of their advisor. | Highly significant |
| Had a mentor other than one's advisor | 54% | Women (62%) were much more likely than men to have had a mentor other than their advisor (48%). | Highly significant |
| Imposter syndrome (scale is 7 to 35) with a higher score meaning respondent feels more like an imposter | 19.2 | Women's scores (19.9) were higher than men's (18.8) meaning women felt more like imposters than men. | Highly Significant |
| Changed advisors while in graduate school | 28% | No | — |
| Respondent limited career options because of someone else | 44% | No | — |
| Respondent was currently a postdoc | 53% | No | — |
| Respondent had completed a postdoc | 21% | No | — |
| Time since degree | 2.6 years | No | — |

**Table 2: Characteristics of Variables in the Study that Affect Attrition**

* Highly significant → *p*-value < 0.01

B. Multivariate analysis

We found two direct effects on working out of astronomy in 2015: working out of astronomy in 2012 and encouragement from one's advisor to attain goals. There were four direct effects on working out of astronomy in 2012: changing advisors, not taking a post-doc, relocating for a spouse or partner, and limiting career options for someone else. Gender is directly related to relocating for a spouse or partner and to encouragement from one's advisor to attain goals. Also, gender is indirectly related to changing advisors and limiting career options for someone else. The only factor affecting working out of astronomy that does not display a gender effect is having been a post-doc. Figure 3 presents the results from the 2015 survey.

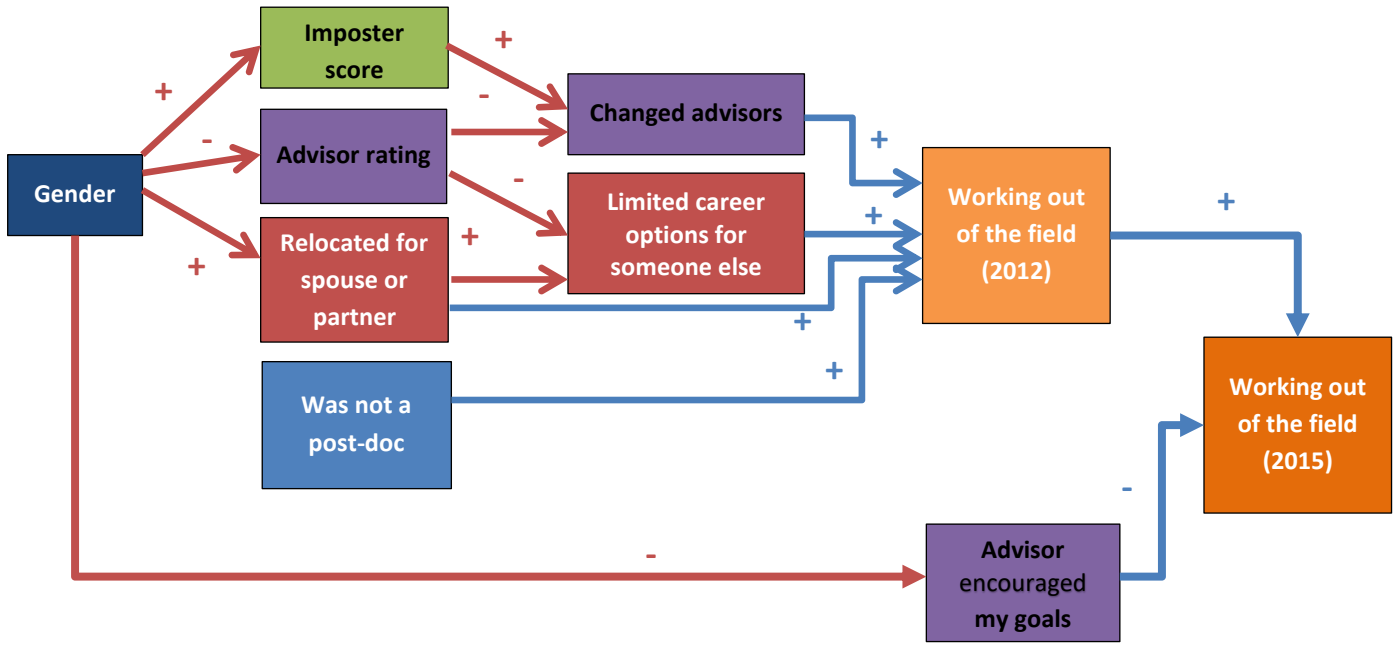

**Figure 3: Direct and Indirect Effects on Working out of the Field of Astronomy in 2015**

In Figure 4, we present the magnitude of the direct effects in 2015. While working out of field in 2012 has the largest impact, we see that changing the perception of the advisor by just one step (from Disagree to Strongly Disagree, for example) means an individual is 1.6 times less likely to be working in astronomy in 2015 – after taking into account whether they were working out of the field in 2012. A two-step change (from Agree to Strongly Disagree, for example) means an individual is 2.6 times less likely to be working in astronomy is 2015. Finally, a drop from Strongly Agree to Strongly Disagree results in being 4.3 times less likely to be working in astronomy in 2015. When these results are considered in light of the responses shown in Figure 2, we see the indirect effect of gender on working out of the field of astronomy.

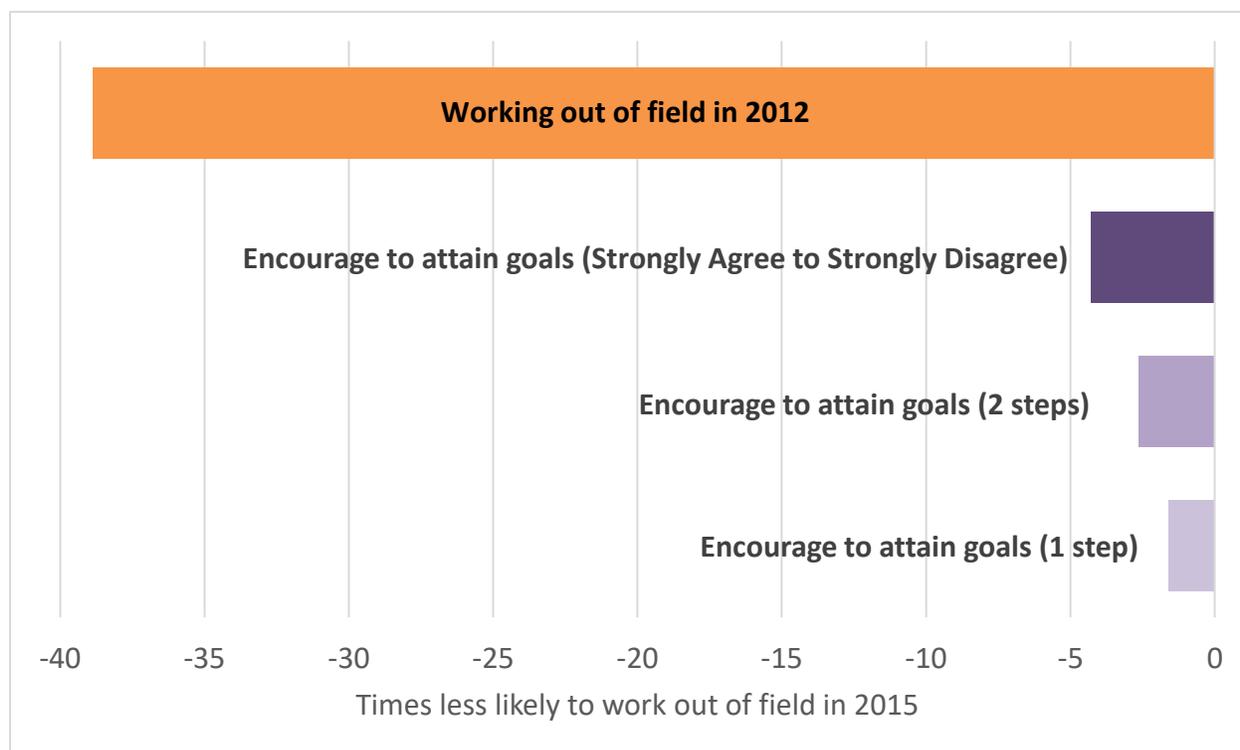

**Figure 4: Magnitude of the Direct Effects on Working Out of Field**

VI. IMPLICATIONS

We find that multiple factors affect persistence in or attrition from astronomy. While gender alone does not directly impact whether one is working in astronomy, gender does affect persistence indirectly. The advising relationship is very important in persistence, both in the doctoral program and in its effects after completion of the doctorate. The literature shows that men and women differ little in the ways in which they choose advisors, and our results show that men are more likely than women to report that their advisors are encouraging them in their careers, giving them credit as appropriate, advocating for them, and supporting their career paths of choice. We need more research on gender differences in the ways that graduate students experience the advising relationship so we can better prepare advisors to work with all students.

In order to reduce attrition from the fields of astronomy and physics, efforts toward improving the advising relationship and mitigating the two-body problem must be made early in the career path. The strongest predictor of working out of field in 2015 was working out of the field in 2012. In other words, once trained astronomers leave, the likelihood of their returning is low.

While LSAGS was designed to measure persistence in the field, persistence is not necessarily the only goal. Satisfaction and use of training are also important career considerations. AIP's other studies have shown that for their first jobs, astronomers and physicists in all employment sectors use their scientific and technical skills on the job and that most feel satisfied with their initial employment outcomes [30].

And while the study was originally designed to examine differences between men and women in attrition from the field, our results point to much larger issues that affect all students.

Our results show that for both men and women, the importance of advisors cannot be underestimated. Even years after graduate school, a good relationship with a doctoral advisor continues to have effects on persistence in the field. Early intervention, in the form of training advisors, could make a great deal of difference in the retention of astronomers after they receive their doctorates.  In the third iteration of LSAGS, we tried to determine the characteristics of advisors that affect persistence in the field. Out of the multiple items that we tested, only one—"my advisor encouraged me to attain my goals"—had a significant effect on persistence in astronomy and physics. This implies that in order to retain students long-term, advisors have a simple task—encouraging students toward the student's own goals. The ways that this can be done would require another study interviewing students in-depth about the specific efforts that advisors made (or did not make) toward this effort.

Training advisors to put the goals of the students foremost in the advisor/student relationship may not be easy, but it is perhaps less complicated that solving the two-body problem, which also has effects on attrition. Nevertheless, our studies document the importance of balancing work life with home life on career outcomes. The two-body problem continues to need solutions both at the individual level and at an organizational level. All efforts that employers, administrators, and institutions can put toward solving the two-body problem and improving the advising relationship should reap benefits in the retention of talented scientists.

The authors appreciate the support of the NSF through grant AST 1347723.